\documentclass[prb,twocolumn]{revtex4-1}%
\usepackage{amsmath}
\usepackage{graphicx}
\usepackage{amsfonts}
\usepackage{amssymb}%
\providecommand{\U}[1]{\protect\rule{.1in}{.1in}}

\begin{document}
\title{Photoinduced melting of superconductivity in the high-T$_{c}$ superconductor
La$_{2-x}$Sr$_{x}$CuO$_{4}$ probed by time-resolved optical and THz techniques.}
\author{M. Beyer,$^{1}$ M. Beck,$^{1}$ D. St\"{a}dter,$^{1}$ H. Sch\"{a}fer,$^{1}$ 
V.V. Kabanov,$^{2,3}$ G. Logvenov,$^{4,5}$ I. Bozovic,$^{4}$ G. Koren,$^{6}$
and J. Demsar$^{1,2,3}$}
\affiliation{$^{1}$Physics Department, Center of Applied Photonics, University of Konstanz,
D-78457 Konstanz, Germany }
\affiliation{$^{2}$Complex Matter Dept., Jozef Stefan Institute, Ljubljana, Slovenia}
\affiliation{$^{3}$Zukunftskolleg, University of Konstanz, D-78457 Konstanz, Germany}
\affiliation{$^{4}$Brookhaven National Laboratory, Upton, NY, USA}
\affiliation{$^{5}$Max-Planck Institute for Solid State Research, D-70569, Stuttgart, Germany}
\affiliation{$^{6}$Physics Department, Technion, Haifa, 32000 Israel}
\date{\today}

\begin{abstract}
Dynamics of depletion and recovery of superconducting state in La$_{2-x}%
$Sr$_{x}$CuO$_{4}$ thin films is investigated utilizing optical pump-probe and
optical pump - THz probe techniques as a function of temperature and
excitation fluence. The absorbed energy density required to suppress
superconductivity is found to be about 8 times higher than the
thermodynamically determined condensation energy density and nearly
temperature independent between 4 and 25 K. These findings indicate that
during the time when superconducting state suppression takes place
($\approx0.7$ ps), a large part (nearly 90\%) of the energy is transferred to
the phonons with energy lower than twice the maximum value of of the SC gap
and only 10 \% is spent on Cooper pair breaking.

\end{abstract}
\maketitle

\section{Introduction}

In the last decade or so numerous real-time studies of carrier relaxation
dynamics in cuprate superconductors have been performed utilizing pump-probe
techniques. The initial studies were aimed at the understanding of
relaxation,\cite{Han,Stevens,Kabanov,
EPL,Segre,Schneider,THzAveritt,KaindlBiSCO,RT} and the interplay between the
superconducting gap and the normal state
pseudogap.\cite{YBCOOver,kaindl,YBCO124,KusarPRB} Recently, however, several
reports explored the dependence on excitation intensity of depletion of
superconducting state\cite{Kusar,Giannetti1,Pashkin,Giannetti2} as well as of
photoinduced structural dynamics\cite{Carbone} and phase
transitions.\cite{Gedik,Radovic}

It has been known since 1971 that an intense laser pulse can destroy the
superconducting (SC) state non-thermally - the absorbed energy density is
lower that the energy density required to heat up the sample to the critical
temperature.\cite{Testardi} The rapid development of stable amplified laser
systems, producing optical pulses with sub 100 fs pulse duration has enabled
studies of dynamics of SC suppression in real time.\cite{MgB2} This technique
enables direct measurement of the SC condensation energy, $E_{c}$ (the
difference in the free energy density between the SC and normal states at zero
temperature). If, following photoexcitation with a fs optical pulse, the
absorbed energy remains in the electronic subsystem during the process of SC
state destruction, the absorbed (optical) energy density required to suppress
SC, $E_{opt}$, should be equal to $E_{c}$. In conventional superconductors
$E_{c}$ can be directly determined by measuring the thermodynamic critical
magnetic field, $B_{c}(0)$, where $E_{c}=B_{c}^{2}(0)/2\mu_{0}$. However, in
high-$T_{c}$ cuprate superconductors the critical magnetic fields are very
high and hardly accessible experimentally. To determine $E_{c}$ in cuprates,
the T-dependence of the electronic specific heat $C_{e}(T)$ has been
studied.\cite{Loram,CpLSCO} However, since $C_{e}(T)$ is determined by
measuring the total specific heat of the SC sample, and subtracting the phonon
part (obtained by measuring the specific heat of an impurity doped non-SC
sample), it may be prone to some uncertainty.

In a recent optical pump-probe (OPP) study, an attempt was made to determine
$E_{c}$ in single crystals of high temperature superconductor La$_{2-x}%
$Sr$_{x}$CuO$_{4}$ (LSCO) by means of ultrafast optics.\cite{Kusar} $E_{opt}$
was found to be about one order of magnitude higher than the thermodynamically
determined $E_{c}$.\cite{CpLSCO} This large difference can hardly be
attributed to the experimental uncertainties, so it was concluded that the
major part of $E_{opt}$ is transferred to the phonon subsystem on the
sub-picosecond timescale.\cite{Kusar} Indeed, a recent optical-pump -- THz
probe (OPTP) study in optimally doped YBa$_{2}$Cu$_{3}$O$_{7-\delta}%
$\ (YBCO),\cite{Pashkin} showed rapid heating of the two infrared-active
c-axis phonon modes on the timescale of 150 fs.\cite{Pashkin} Moreover,
$E_{opt}$ was also found to be about 10 times larger than $E_{c}%
$.\cite{Pashkin} Similar studies on Bi$_{2}$Sr$_{2}$CaCu$_{2}$O$_{8+\delta}$
(Bi2212) single crystals \cite{Giannetti1,Giannetti2,Uwe} also indicate that
$E_{opt}\approx10E_{c}$.

Here we report the results of study of La$_{2-x}$Sr$_{x}$CuO$_{4}$ (x = 0.08,
0.16, 0.21) thin films using both OPP and OPTP. In addition to the
low-temperature measurements, the temperature dependence of $E_{opt}$ was
studied. We find that, within the experimental uncertainty, both OPP and OPTP
give an identical value of $E_{opt}$ for LSCO. This observation is of
particular importance, since OPTP is probing the gap resonantly. The obtained
value of $E_{opt}$ is nearly identical to the one extracted from the OPP
studies on LSCO single crystals.\cite{Kusar} In addition, we find almost no
temperature dependence of $E_{opt}$ below $T_{c}$. This observation, together
with the fact that $E_{opt}\approx8E_{c}$, strongly suggest that during the
time when the SC state suppression takes place most of the absorbed energy is
transferred to phonons with energy lower than $2\Delta$, where $\Delta$
denotes the value of the SC gap maximum. Because of strong reduction in the
density of states at low energies in d-wave superconductors, Cooper pair
breaking by phonons with $\hbar\omega<2\Delta$ is strongly suppressed. In LSCO
about $90$ $\%$ of the absorbed energy is directly released to the lattice due
to strong \textit{e-ph} relaxation and only $\sim10$ $\%$ is spent on
destruction of the condensate (Cooper pair breaking).

\section{Experimental}

The LSCO thin films used in this study were grown on LSAO substrate either by
molecular beam epitaxy (MBE)\cite{MBE} or by pulsed laser deposition
(PLD)\cite{PLD}. The film grown by MBE (x = 0.16) had a thickness of 52 nm
with surface roughness less than one monolayer, and exhibited a critical
temperature, T$_{c}$, of 31 K. The films grown by PLD (x = 0.08, 0.21) were
$75\pm5$ nm thich with surface roughness of about 2 nm, and Tc's of 22, and 25
K, respectively. The OPP experiments were performed in a high sensitivity
pump-probe set-up utilizing a 250 kHz regenerative Ti:Sapphire amplifier,
delivering 50 fs pulses at 800 nm (1.55 eV), and a fast-scan
technique.\cite{BBHanjo} The pump and probe beam diameters were measured
accurately with a CCD camera, and were 120 $\mu$m and 60 $\mu$m, respectively,
to insure a homogeneous excitation profile. The OPTP experiments were
performed using a set-up based on the same amplifier system, employing a
large-area photoconductive finger emitter generating phase-locked THz pulses
with spectrum covering 0.2 - 3 THz range.\cite{OptExp} The THz beam was
focussed to about 1.5 mm, while the pump beam diameter was about 3 mm, to
insure a homogeneous excitation profile. Experiments were conducted in a wide
range of excitation fluences (3 orders of magnitude), spanning from $\emph{F}$
= 0.1 $\mu$J/cm$^{2}$ up to 200 (100) $\mu$J/cm$^{2}$ in the OPP (OPTP) configurations.

\section{Results}

Figure 1 presents the photoinduced (PI) dynamics probed by the OPP and OPTP
configuration at 4K. The PI reflectivity ($\Delta R/R$) traces for different
excitation fluences (in $\mu$J/cm$^{2}$) are shown in Fig. 1a. Like in many
cuprate superconductors,\cite{EPL,YBCO124,KusarPRB} two distinct relaxation
components are observed in time-domain studies in LSCO, one (A) being present
only below the SC critical temperature, while the other one (B) being
sensitive to the opening of the normal state pseudogap.\cite{KusarPRB,Kusar}
At low temperatures and excitation densities the component A is
dominant.\cite{KusarPRB} It is characterized by a rise-time of $\approx$ 0.7
ps and the decay time on the 10 ps timescale. The component B, on the other
hand, is characterized by a sub 100 fs rise-time and the recovery that is
weakly dependent on temperature and excitation fluence, with the recovery
time-scale not exceeding 1 ps. Upon increasing the excitation fluence,
$\emph{F}$, the component A shows saturation (see Fig. 2a), while the
component B increases linearly with $\emph{F}$ up to much higher fluences.
Given the fact that the component A can be attributed to suppression and
recovery of the superconducting gap, its saturation observed at high
excitation densities can be naturally attributed to the complete destruction
of superconductivity.

To measure the dynamics of SC suppression and recovery in the THz range, over
large range of $\emph{F}$, we have performed studies of the spectrally
integrated conductivity change (often refereed to as a 1-dimensional scan
technique).\cite{Future} Here the photoinduced change in the optical
conductivity, $\Delta\sigma$, is proportional to the photoinduced change in
the transmitted electric field $\Delta E_{tr}(t^{\prime}=t_{0})$, where
$t_{0}$ is a fixed point of $E_{tr}(t^{\prime})$ trace - see the insert to
Fig. 1b. In LSCO $2\Delta$ $\gg1$ THz, so the SC-induced change in the THz
optical conductivity modifies the transmitted electric field transient mainly
because of the appearance of the so-called kinetic inductance.\cite{MgB2} To
achieve a high dynamic range, $\Delta E_{tr}(t_{0}=-0.35$ ps$)$ was recorded
(marked by an arrow in insert to Fig. 1b), where the change in the electric
field transmitted through the sample, $\Delta E_{tr}$, corresponding to the
transition between the SC and normal states, is the highest. The PI traces
recorded in near optimally doped sample (La$_{1.84}$Sr$_{0.16}$CuO$_{4}$) at 4
K and at various fluences are displayed in Fig. 1b. Similarly to the component
A from the OPP data, the amplitude of the induced change initially increases
linearly with $\emph{F}$\ and shows saturation at high $\emph{F}$, as shown in
Fig. 2b.%

\begin{figure}
[ptb]
\begin{center}
\includegraphics[
height=11.451cm,
width=8.5009cm
]%
{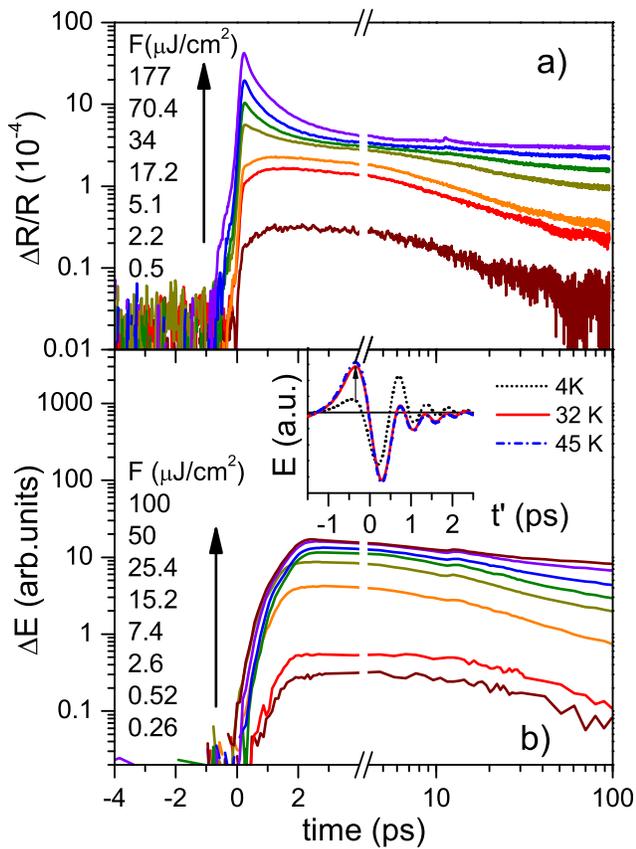}%
\caption{(color online) a) PI reflectivity dynamics in La$_{1.84}$Sr$_{0.16}%
$CuO$_{4}$ at 4 K as a function of excitation fluence. b) The corresponding
dynamics of the PI change in the transmitted THz electric field. Insert: The
THz electric field transient transmitted through the LSCO film on LSAO
substrate at different temperatures. The arrow marks the $t^{\prime}$ where
the largest change between the SC and normal state is observed. The transients
were shifted to 0 ps for display.}%
\end{center}
\end{figure}

Another noteworthy feature of the data shown in Fig. 1 is the dependence of
the recovery dynamics on fluence. At lowest excitation densities, the PI
changes in reflectivity and in THz conductivity almost completely recover,
with the characteristic decay time of $\approx10$ ps (using the exponential
decay fit). At higher excitation fluences, after the initial recovery on the
10 ps timescale, the induced change reaches plateau, with further recovery
proceeding on a much longer timescale. As we discuss below, this plateau can
be attributed to an overall increase in the film temperature and its recovery
is governed by the heat diffusion to the substrate. Indeed, as the excitation
density is increased above $\approx50$ $\mu$J/cm$^{2}$, the PI signal at the
time delay of 100 ps, which is much longer than the characteristic SC state
recovery time of $\approx10$ ps, also shows a saturation behavior - see Figs.
2c and 2d.%

\begin{figure}
[ptb]
\begin{center}
\includegraphics[
height=7.6091cm,
width=8.4987cm
]%
{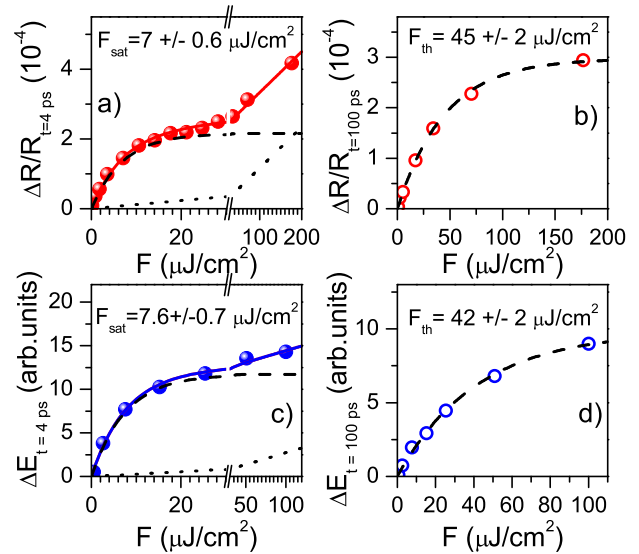}%
\caption{(color online) The evolution of the PI change in reflectivity at time
delay a) 4ps and b) 100 ps as a function of excitation density obtained at 4K
in the OPP configuration. The corresponding evolutions of the PI change in the
THz electric field at $t^{^{\prime}}$ obtained in the OPTP experiment are
shown in c) and d). The data can be well fit with the simple saturation model
(solid line, see text); long dashed and short dashed lines represent the first
and second term, respectively, of the the fit model. The fluence required to
transiently suppress superconductivity is given by $F_{sat}$, while $F_{th}$
matches well the fluence required to thermally heat up the sample to above
T$_{c}$. }%
\end{center}
\end{figure}

In Fig. 2a we plot the $\emph{F}$-dependence of the magnitude of component A,
which is related to SC state suppression. In order to avoid picking up the
contribution from component B, which starts to dominate once the component A
saturates, we plot the magnitude of the signal at the time delay of 4 ps. To
determine the characteristic fluence required to suppress SC, we use a simple
saturation model:
\begin{equation}
\Delta R/R(4ps)=C(1-\exp(-\emph{F}/\emph{F}_{sat}))+D\emph{F}\text{.}
\label{EqSat}%
\end{equation}
The first term corresponds to the saturating part, where $\emph{F}_{sat}$ is
the excitation fluence required to suppress the SC. The second term in
Eq.(\ref{EqSat}) is accounting for small contribution from the component B,
which at high $\emph{F}$ is not negligible despite the fact that the recovery
time of component B is less than 1 ps. $C$ and $D$ are constants. From the
best fit to the data (solid curve in Fig. 2a) we obtain $\emph{F}_{sat}%
=7\pm0.6$ $\mu$J/cm$^{2}$. In Fig. 2c, we plot the $\emph{F}$-dependence of
the corresponding induced change in the transmitted THz electric field,
$\Delta E_{tr}(t_{0})$, again recorded at 4 ps after photoexcitation.
Similarly to the OPP data, the saturation of the induced change is observed
with $\emph{F}_{sat}=7.6\pm0.7$ $\mu$J/cm$^{2}$. Within the uncertainty in the
absolute excitation densities in the two configurations, the values for
$\emph{F}_{sat}$ obtained in the two experimental configurations are identical.

To accurately determine the absorbed energy density, which corresponds to the
optically induced suppression of the SC state, we have measured the dielectric
constants of La$_{2-x}$Sr$_{x}$CuO$_{4}$ at the excitation photon energy of
1.55 eV (800 nm). By measuring the reflectivity R and transmission T through
the film on substrate and through the bare substrate, and using the
appropriate Fresnel equations we numerically solved a system of equations for
R and T.\cite{Harbecke} The extracted complex refractive index of La$_{1.84}%
$Sr$_{0.16}$CuO$_{4}$ is $\widetilde{n}(800$ nm$)=n+ik=2.06+i0.38$. From the
measured reflectivity and the extinction coefficient ($\alpha=5.8\cdot10^{4}%
$cm$^{-1}$) we obtained the\ absorbed energy density that corresponds to
$\emph{F}_{sat}$, $E_{opt}$. For near optimally doped La$_{1.84}$Sr$_{0.16}%
$CuO$_{4}$ $E_{opt}\approx0.35$ Jcm$^{-3}$, which corresponds to $2.4$ k$_{B}%
$K per Cu atom (i.e. $E_{opt}\approx2.4$ K/Cu).

In Figs. 2c and 2d, we plot the $\emph{F}$-dependence of the PI change in
reflectivity and in the transmitted THz electric field, at the time delay of
100 ps. Since this is substantially longer than the timescale for SC recovery,
we can assume that 100 ps after photoexcitation the film is in the
quasi-equilibrium, where the electronic system and the underlying lattice are
thermalized at a given temperature. Similarly to the PI amplitude at short
time delays, the saturation of the signal is observed with a characteristic
fluence $\emph{F}_{th}\approx40-45$ $\mu$J/cm$^{2}$, which corresponds to the
absorbed energy density $E_{th}$ $\approx2$ Jcm$^{-3}=13.7$ K/Cu.

We should note that in our case the excitation density is nearly homogeneous
throughout the probed volume, since i) the pump spot diameter is twice bigger
than the probe spot diameter and ii) the film thickness (52 or 75 nm) is
substantially lower than the optical penetration depth (e.g. $l_{opt}\left(
\text{La}_{1.84}\text{Sr}_{0.16}\text{CuO}_{4}\right)  =170$ nm). Thus the
values for $E_{opt}$ and $E_{th}$ are quite precise.

The absorbed energy density corresponding to transiently suppressing the SC
state, $E_{opt}$ is about 8 times higher than $E_{c}\approx0.3$
K/Cu.\cite{CpLSCO} On the other hand, $E_{opt}$ is by about a factor of 5-6
lower than the energy required to heat up the excited sample volume to above
$T_{c}$. Using the reported data on the total specific heat for the optimally
doped LSCO,\cite{Junod} we obtain $U_{th}=%
{\textstyle\int\nolimits_{4K}^{T_{c}}}
C_{p}(T)dT\approx1.6$ Jcm$^{-3}\approx11$ K/Cu. This value is in very good
agreement with $E_{th}$ $\approx13.7$ K/Cu, implying that at fluences above
$\emph{F}_{th}$, superconductivity is also thermally suppressed, with its
recovery proceeding on the timescale determined by the heat diffusion. The
excellent agreement between $E_{opt}$ obtained by OPP and OPTP techniques, as
well as the agreement between $E_{th}$ and $U_{th}$ strongly indicate that
$E_{opt}\approx8E_{c}$, where $E_{c}$ is determined thermodynamically.%

\begin{figure}
[ptb]
\begin{center}
\includegraphics[
height=6.6316cm,
width=8.5009cm
]%
{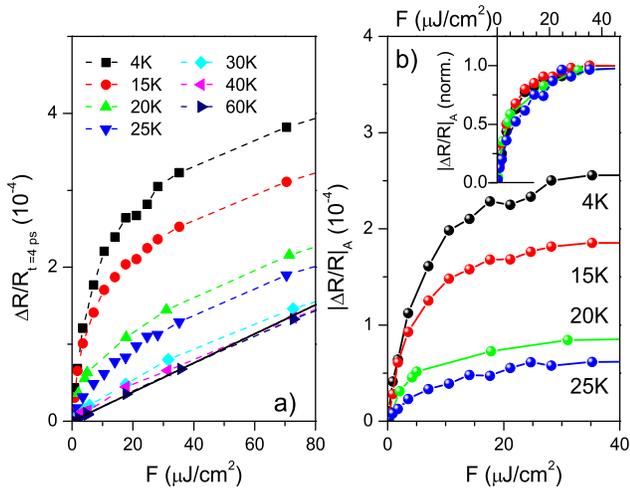}%
\caption{(color online) a) The fluence dependence of the PI reflectivity
change at 4 ps time delay recorded in La$_{1.84}$Sr$_{0.16}$CuO$_{4}$ at
different temperatures. From T$_{c}=31$ K to 60 K all the data fall on the
single curve, showing linear fluence dependence of the signal within the range
of fluences studied. By subtracting this temperature independent linear term
(steming from component B) from the data recorded at different T we obtain the
F-dependence of the component A - shown in panel b). The threshold fluence
$\mathcal{F}_{sat}$ is found to be nearly T-independent, as demonstrated by
normalizing all the curves - see insert.}%
\end{center}
\end{figure}

To gain further insight into the energetics of the photoinduced SC to normal
phase transition, we have performed the first temperature dependent studies of
photoinduced quenching of superconductivity. Importantly, continuous heating
of the sample is in the case of thin films substantially reduced in comparison
to single crystals, since the low temperature thermal conductivity of LSAO
substrate \cite{LSAOthermal} is much higher than that of LSCO (especially in
the c-direction).\cite{LSCOKappa} Therefore the cumulative temperature
increase of the probed spot is negligible enabling such studies all the way up
to close vicinity of T$_{c}$. In Fig. 3, we present the \emph{F}-dependence of
PI reflectivity change at time delay of 4 ps, recorded in near optimally doped
LSCO at several temperatures below and above T$_{c}$ = 31 K. Above T$_{c}$,
all the data fall on the same curve, displaying linear \emph{F}-dependence of
the normal state response. Below T$_{c}$, the signal clearly shows two
contributions, one (A) showing saturation above \emph{F}$_{sat}$, while the
other (B) showing linear \emph{F}-dependence, with the slope being
T-independent in the range of temperatures studied. By subtracting the
contribution of the component B from all the data, we obtain the fluence
dependence of the component A - see Fig. 3b. Interestingly, \emph{F}$_{sat}$
($E_{opt}$) is found to be nearly temperature independent up to 25 K.

Similar data to the one on a near optimally doped film (Fig. 3) were obtained
also in an underdoped x=0.08 and an overdoped x=0.21 films prepared by pulsed
laser deposition. Figure 4 summarizes the results on a x = 0.08 sample in the
optical pump-probe configuration (similar data are obtained in the optical
pump THz probe configuration). Here also, in the normal state the photoinduced
reflectivity change varies linearly with fluence (component B). When the
normal state response is subtracted from the data, the superconducting
response (component A) shows clear saturation (Fig. 4b), with the threshold
fluence, \emph{F}$_{sat}$, required to transiently suppress superconductivity
being nearly temperature independent all the way to T$_{c}=22$ K.%

\begin{figure}
[ptb]
\begin{center}
\includegraphics[
height=6.4932cm,
width=8.5536cm
]%
{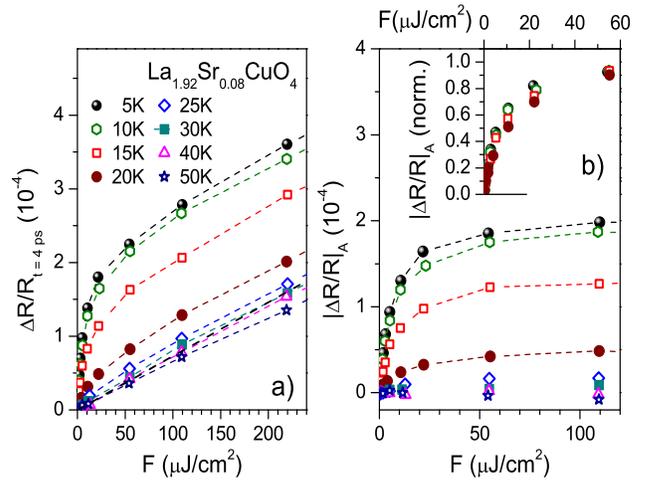}%
\caption{(color online) a) The evolution of the PI change in reflectivity at a
time delay of 4ps in La$_{1.92}$Sr$_{0.08}$CuO$_{4}$ film recorded at
different temperatures. As in the case of the near optimally doped film
(x=0.16), above T$_{c}\ $($24$ K) all the data fall on the single curve,
showing linear fluence dependence (due to component B). b) When the linear in
\emph{F} contribution is subtracted from the data, only the part showing
saturation (component A) remains. Insert shows the normalized \emph{F}-
dependence of component A, demonstrating the absence of T-temperature of
$\mathcal{F}_{sat}$ ($E_{opt}$).}%
\end{center}
\end{figure}

Figure 5 summarizes the values of the $E_{opt}$ for various dopings obtained
in La$_{2-x}$Sr$_{x}$CuO$_{4}$ thin films together with the values obtained on
the single crystals.\cite{Kusar} For all films the values of the complex
refractive index were determined by measuring reflectivity and transmission at
near normal incidence through the film and the bare substrate and numerical
analysis of Fresnel equations.\cite{Harbecke} The larger error bars on films
grown by pulsed laser deposition stem from the uncertainty in the film
thickness of $\pm5$ nm.%

\begin{figure}
[ptb]
\begin{center}
\includegraphics[
height=7.1852cm,
width=8.4526cm
]%
{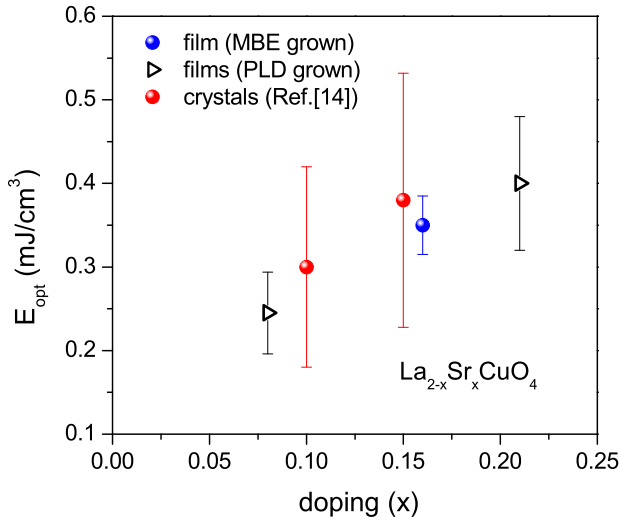}%
\caption{(color online) Doping dependence of $E_{opt}$ in La$_{2-x}$Sr$_{x}%
$CuO$_{4}$, combining the results obtained on thin films with those obtained
in single crystals.\cite{Kusar}}%
\end{center}
\end{figure}

\section{Discussion}

The observation of $E_{opt}$ being nearly T-independent below $T_{c}$ with
$E_{opt}/E_{c}\approx8$ is in striking contrast to similar studies on the NbN
superconductor.\cite{Beck} There, $E_{opt}$ was found\cite{Beck} to be equal
to $E_{c}$, with both following the temperature dependence of $\Delta^{2}$.
The large disproportionality between the $E_{opt}$ and $E_{c}$ was
argued\cite{Kusar} to be due to transfer of a large fraction of the absorbed
optical energy to high frequency ($\hbar\omega>2\Delta$) phonons on the
timescale required to suppress superconductivity ($\approx0.7$ ps in LSCO). On
the other hand, in conventional superconductors like MgB$_{2}$\cite{MgB2} and
NbN,\cite{Beck} the optically induced suppression of SC state also takes place
via a two-step process.\cite{MgB2} Here, high energy electrons(holes) created
by absorption of photons with energy much larger that the gap energy, first
relax towards the gap via \textit{e-e} and \textit{e-ph} scattering. The fact
that the pair-breaking time ($\approx10$ ps) is large and excitation density
dependent\cite{MgB2} was attributed to the fact, that during the initial
\textit{e-e} and \textit{e-ph} scattering a high density of optical phonons
with energies larger than $2\Delta$ is generated, which subsequently break
Cooper pairs\cite{RT} on the 10 ps timescale. The question arises, then, why
$E_{opt}/E_{c}\approx8$ in LSCO and YBCO, while $E_{opt}\simeq E_{c}$ in NbN?
To answer this question, we discuss the nature of \textit{e-ph} scattering
processes and compare the phonon spectra and sizes of the superconducting gaps
$\Delta$ between the two cases.

As it was pointed out also in Ref.\cite{kabalex}, the dominant process in
cooling of photoexcited electrons (holes) is the inelastic scattering by
phonons; \textit{e-e} scattering is dominant only if the energy of the
electron (hole) is far from the Fermi energy. Here we would like to address
the issue of the energy and momentum distributions of phonons generated by hot
electrons (holes). Generally, the transition rate of the electron with
momentum ${\mathbf{k}}$ and energy $\epsilon_{\mathbf{k}}=\hbar^{2}k^{2}/2m$
to the state with the momentum ${\mathbf{k}^{^{\prime}}=\mathbf{k}+\mathbf{q}%
}$ via emission of a phonon with the momentum ${\mathbf{q}}$ and frequency
$\omega$ is given by
\begin{equation}
w_{q}=\frac{2\pi}{\hbar}|M_{\mathbf{q}}|^{2}(N_{\mathbf{q}}+1)(1-n(\epsilon
_{\mathbf{k}-\mathbf{q}}))\delta(\epsilon_{\mathbf{k}}-\epsilon_{\mathbf{k}%
-\mathbf{q}}-\omega), \label{Eq1}%
\end{equation}
where $M_{\mathbf{q}}$ is the matrix element which depends on the
\textit{e-ph} scattering mechanism while $N_{\mathbf{q}}$ and $n(\epsilon
_{\mathbf{k}-\mathbf{q}})$ are the equilibrium distribution functions of
phonons and electrons, respectively. By integrating Eq.(\ref{Eq1}) over
${\frac{d^{3}q}{{(2\pi)^{3}}}}$ one obtains the number of phonons generated by
a hot electron (or hole) per unit of time. Note that this expression
represents only one part of the \textit{e-ph} collision
integral.\cite{kabalex} The second part describes the phonon absorption
(mainly Cooper pair breaking), which is strongly energy dependent in the case
of a superconductor with the gap in the quasiparticle excitation spectrum.
Frequently, one is not interested in the momentum distribution of phonons and
electrons and considers their distribution as a function of energy only.
However, here one should consider also the momentum distribution of generated
phonons. To find the rate of generation of phonons at a particular wave-vector
$q=|\mathbf{q}|$, we integrate the expression (\ref{Eq1}) over the angular
part of $\mathbf{q}$, keeping the modulus of $q$ constant. Since experiments
are performed at low temperatures and rather low excitation densities, we can
assume that $N_{\mathbf{q}}\approx n(\epsilon_{\mathbf{k}-\mathbf{q}}%
)\approx0$. Under this assumption, we obtain
\begin{equation}
w_{q}={\frac{(2\pi)^{2}|M_{\mathbf{q}}|^{2}m}{\hbar^{3}{kq}},} \label{eq2}%
\end{equation}
where $m$ is the effective mass of charge carriers. Eq.(\ref{eq2}) describes
the phonon generation rate by hot electrons (holes). For metals, we can
substitute $k$ in Eq.(\ref{eq2}) by the Fermi wave vector $k_{F}$ assuming\ a
large Fermi energy. Because of the energy conservation, $\hbar^{2}{\mathbf{k}%
}^{2}/2m-\hbar^{2}({\mathbf{k}}-{\mathbf{q}})^{2}/2m=\hbar\omega$, and
momentum conservation, the phonon wavevector in Eq.(\ref{eq2}) is restricted
by $q_{min}<q<q_{max}$ where $q_{min}\approx m\omega/\hbar k_{F}$ and
$q_{max}\approx2k_{F}$.

In most metals, screening is strong, and the matrix element of the electron
interaction with optical phonons (deformation optical phonon scattering, DO)
is\cite{gantmah} momentum independent $|M_{q}^{DO}|^{2}=const$. Since
$w_{q}^{DO}\propto1/q$ - see Eq.(\ref{eq2}) - generation of optical phonons at
$q\approx q_{min}$ is dominant. Yet, since the phase volume of $q\approx
q_{min}$ phonons is small, when we integrate Eq.(\ref{eq2}) with
$q^{2}dq/(2\pi)^{3}$, the total energy accumulated by the low-$q$ optical
phonons remains small.

The matrix element of the interaction of electrons with acoustic phonons
(deformation acoustic phonon scattering, DA) has $|M_{q}^{DA}|^{2}\propto q$
dependence,\cite{gantmah} and thus $w_{q}^{DA}$ is momentum independent;
hence, the energy is mainly transferred to the high-$q$ (high-$\omega$)
acoustic phonons due to their large phase volume.

In NbN, the phonon spectrum consists of acoustic branches extending up to
$\approx28$ meV and the weakly dispersing optical branches at $\approx60$
meV,\cite{NbNneutrons} while $2\Delta=11$ meV \cite{Dressel}. From the above
considerations, it follows that the inital \textit{e-ph} scattering process
generates a high density of optical phonons and high-frequency acoustic
phonons, all of which have energy larger than $2\Delta$. All of these phonons
can effectively break Cooper pairs. When on the timescale of Cooper pair
breaking ($\approx10$ ps in NbN) the quasi-equilibrium is established between
the populations of quasiparticles and phonons with $\hbar\omega>2\Delta$, the
excess energy is almost exclusively stored in the electronic
subsystem.\cite{RT,comment} Therefore it is no surprise that $E_{opt}\simeq
E_{c}$, as experimentally observed.

In high-$T_{c}$ superconductors, however, in addition to the DO and DA
processes, the scattering on the c-axis polar (infrared active) optical
phonons (polar optical scattering, PO) is possible. Since the $c$-axis plasma
frequency is low, the polar optical phonons are unscreened.\cite{caxisphonons}
As it is well known,\cite{gantmah} the matrix element of the interaction of
electrons with polar optical phonons scales as $|M_{q}^{PO}|^{2}\sim1/q^{2}$.
Since $w_{q}^{PO}\propto1/q^{3}$, the non-equilibrium distribution function of
polar c-axis phonons is strongly peaked at low-$q$. Moreover, in the case of
PO, the integration over the phase volume does not cancel the $1/q^{3}$
dependence of the generation rate. Therefore for scattering on polar optical
phonons also most of the energy is accumulated near the $\Gamma$ point of the
Brillouin zone, and it can be sizeable $-$ as demonstrated.\cite{Pashkin}

In LSCO, acoustic branches extend up to $\approx10$ meV, while the spectrum of
optical phonons continuously extends all the way to 100 meV with a maximum in
the phonon density of states near 20-30 meV.\cite{PDOSlsco} On the other hand,
in LSCO\ the value of the gap maximum in the anti-nodal direction
is\cite{LSCOgap} $\Delta\approx15$ meV. Comparison of $2\Delta\approx30$ meV
with the phonon density of states, taking into account the different
\textit{e-ph} scattering mechanisms, reveals that in LSCO the \textit{e-ph}
scattering creates a high density of phonons with $\hbar\omega$ $<2\Delta$.
The Cooper pair breaking process by absorption of phonons with $\hbar\omega$
$<2\Delta$ is strongly suppressed even in d-wave superconductors, due to the
energy and momentum conservation laws and the strong reduction in the density
of states at low energies \cite{Hirschfeld}. Therefore the generation of
phonons with $\hbar\omega$ $<2\Delta$ can present a parallel energy relaxation
channel, competing with Cooper pair breaking. Since $\Delta(T)$ does not
change substantially between 4K and 0.8 T$_{c}$, it is conceivable that
relaxation via these channels is the cause of the observed large difference
between $E_{opt}$ and $E_{c}$.

The calculation of phonon emission rates for various possible \textit{e-ph}
scattering processes is generally difficult and clearly beyond the scope of
this paper. However, for the generation of c-axis polar modes via the polar
optical phonon scattering such a calculation is rather straightforward, and
can be directly compared with the recent studies of time-resolved c-axis THz
conductivity dynamics in the superconducting state of YBCO.\cite{Pashkin} In
this work in addition to the QP relaxation dynamics the dynamics of two
infrared active phonons have been investigated, showing a remarkably fast
increase in the phonon population density of the apical phonon on the
timescale of $\approx150$ fs.\cite{Pashkin} Integrating Eq.(\ref{eq2}) with
$q^{2}dq/(2\pi)^{3}$ and taking into account that $|M_{q}^{PO}|^{2}=\frac
{1}{4\pi\varepsilon_{0}}{\frac{2\pi e^{2}\omega}{{\kappa q^{2}}}}$ where
$\kappa^{-1}=\epsilon_{\infty}^{-1}-\epsilon_{0}^{-1}$, and $\epsilon
_{0},\epsilon_{\infty}$ are the static and the high frequency dielectric
functions, respectively, we obtain (see also Eq. (4.53) in Ref.\cite{gantmah}%
)
\begin{equation}
\tau^{-1}=\frac{1}{4\pi\varepsilon_{0}}{\frac{me^{2}\omega}{\hbar^{2}{\kappa
k_{F}}}}\ln{(q_{max}/q_{min}),} \label{Eq.3}%
\end{equation}
where $\varepsilon_{0}$ is the permittivity of vacuum. Using the values for
the c-axis dielectric constants $\epsilon_{0}\approx30$, $\epsilon_{\infty
}\approx5,$\cite{dielectric} $k_{F}\sim\pi/a$ where $a$ is the lattice
constant, and the free electron mass, we obtain $\tau\sim5$ fs for the
generation of one c-axis polar optical phonon with $\hbar\omega=50$ meV.
Therefore, in the absence of phonon re-absorption processes, the electron
(hole) at $\epsilon=1$ eV above(below) the Fermi energy releases its excess
energy to c-axis polar optical phonons on the timescale of $\tau\epsilon
/\hbar\omega=100$ fs, consistent with measurements on YBCO.\cite{Pashkin}
Indeed, the maximum photoinduced softening of the apical phonon in YBCO is
comparable to the effect induced by a thermal phonon population at
$T\approx200$ K,\cite{Pashkin} suggesting that the phonon distribution
function is on the sub picosecond timescale highly non-thermal.

As we have shown, following absorption of high energy photons, hot electrons
(holes) in cuprate superconductors rapidly relax towards the gap energy by
generating large densities of phonons (predominantly zone edge acoustic and
zone center polar optical phonons). Since in cuprates (investigated to date)
$2\Delta$ is comparable to the phonon cut-off frequency, large portion of the
absorbed energy is on the 100 fs timescale transferred to $\hbar\omega$
$<2\Delta$ phonons without affecting superconductivity.

The doping dependent study - see Fig. 5 - reveals a nearly linear increase of
$E_{opt}$ as a function of doping. Qualitatively, the increase of $E_{opt}$
from the strongly underdoped LSCO towards optimal doping could be understood
within the above scenario to be a result of an increase in the gap energy
scale and therefore the number of phonon modes with $\hbar\omega$ $<2\Delta$
upon doping. However, given the fact that $\Delta$ shows a decrease upon
entering the overdoped range,\cite{Yuli,TcVsDop} one would expect $E_{opt}$ to
follow this dependence. The fact that $E_{opt}$ shows no decrease in the
overdoped regime, however,\ implies that the generation rate of $\hbar\omega$
$<2\Delta$ phonons is also affected by doping, increasing with $x$.

The question arises, however, whether the observation that the generation rate
of $\hbar\omega$ $<2\Delta$ phonons is increasing with doping necessary means
that the electron-phonon coupling constant $\lambda$ is increasing with
doping? The short answer is no. The dimensionless e-ph coupling constant
$\lambda$, which in the BCS theory determines the value of the critical
temperature, is defined as
\begin{equation}
\lambda=2%
{\textstyle\int\limits_{0}^{\infty}}
d\omega\frac{\alpha^{2}F\left(  \omega\right)  }{\omega},
\end{equation}
where $\alpha^{2}F\left(  \omega\right)  $ is the Eliashberg function, which
depends only on the phonon frequency $\omega$. In general, however, the
Eliashberg function depends also on the energy of electron. For simplicity,
let us consider good metals, where screening is large, and the momentum
dependence of electron-phonon interaction becomes less important. In this
case, according to the Eq.(28) of Ref.[36], the relaxation of hot electrons
can be described in terms of the Eliashberg function $\alpha^{2}F(\omega,\xi
)$, which depends both on phonon frequency as well as the electron energy,
$\xi$, counted from Fermy energy, $E_{F}$.\cite{kabalex,Allen} Since the
characteristic energy of electron in the thermodynamic equilibrium is small
(of the order of $k_{B}T\sim k_{B}T_{c}<<E_{F}$), it is usually neglected in
the Eliashberg function which enters into Eq.(5), i.e. $\alpha^{2}F(\omega
,\xi\approx0)=\alpha^{2}F(\omega)$. However, after photo excitation the
characteristic energy of hot electrons is sufficiently higher then $\omega$.
Since the phonon generation rate by hot electrons is proportional to the
spectral function of the electron-phonon interaction,\cite{kabalex} neglecting
the $\xi$ dependence in the Eliashberg function may be incorrect. If the
function $\alpha^{2}F(\omega,\xi)$ at $\xi$ of the order of the photon energy,
$\xi_{ph}$, is sufficiently different from Eliashberg function at $\xi=0$,
this can have a very strong effect on the optical energy required for the
depletion of the superconducting state. For example, if hot electrons interact
mainly with the low frequency phonons (i.e. $\alpha^{2}F(\omega,\xi_{ph})$ is
peaked at low frequencies), while $\alpha^{2}F(\omega,\xi=0)$ is peaked in the
range of high-frequency phonons, the initial relaxation of hot electrons may
create phonons which do not interact with electrons at $\xi\sim T$ and do not
give rise to Cooper pair breaking. In the oposite case, when hot electrons
interact mainly with the high frequency phonons, while $\alpha^{2}F(\omega
,\xi=0)$ is peaked at low frequencies, the electrons near the Fermi energy are
only weakly coupled to the high frequency optical modes. In such a case, the
optical phonons first need to decay into the low-frequency ones before Copper
pair-breaking can take place, and the pair-breaking process will be strongly delayed.

From the experiments performed on cuprates thus
far,\cite{Kusar,Pashkin,Giannetti1,Giannetti2} the characteristic time-scales
for suppression of superconductivity are in the range of 100 to several 100
femtoseconds. From the above considerations we suggest, that\ in cuprates the
dominant part of the absorbed energy is on the 100 fs timescale transferred to
$\hbar\omega$ $<2\Delta$ phonons, which do not take part in the Cooper
pair-breaking process. To further clarify these interesting observations,
systematic studies of the Cooper pair-breaking process as a function of
excitation photon energy, measuring both the dynamics of the condensate as
well as that of the phonons \cite{Pashkin} are required, some of which are
already on the way.

\section{Conclusions}

Utilizing optical pump-probe and optical pump-THz probe techniques, we have
performed systematic studies of the dynamics of suppression and recovery of
superconducting state in La$_{2-x}$Sr$_{x}$CuO$_{4}$ thin films. The results
clearly demonstrate that the absorbed energy density required to suppress
superconductivity exceeds the thermodynamically determined condensation energy
by a factor of $\approx8$, yet this energy is by a factor of $\ 5-6$ lower
than the energy required to thermally heat up the photoexcited volume to above
T$_{c}$. While there is some uncertainty in the existing determination of the
condensation energy, it could hardly explain the observed high $E_{opt}/E_{c}$
ratio. Moreover, the lack of\ temperature dependence of $E_{opt}$ up to
$\approx0.8$ T$_{c}$ clearly suggests that the measured $E_{opt}>E_{c}$, and
suggests the existence of a parallel energy relaxation channel. By considering
various \textit{e-ph} scattering mechanisms we conclude that in cuprates a
large amount of the absorbed optical energy is transferred to phonons with
$\hbar\omega$ $<2\Delta$, on the 100 fs timescale. Since Cooper pair breaking
by phonons with $\hbar\omega$ $<2\Delta$ is strongly suppressed, it is the
\textit{e-ph} scattering to these phonons that presents the parallel
relaxation path.

\begin{acknowledgments}
J.D. acknowledges discussions with T. Dekorsy, A. Leitenstorfer, A. Pashkin,
K.W. Kim and P. Leiderer. The research was supported in parts by the
Sofja-Kovalevskaja Grant from the Alexander von Humboldt Foundation,
Zukunftskolleg and CAP at the University of Konstanz, joint German-Israeli DIP
Project - grant No. 563363, and the Karl Stoll Chair in advanced materials
(Technion).The work at BNL was supported by US DOE contract MA-509-MACA.
\end{acknowledgments}


\begin{thebibliography}{99}
\bibitem {Future}Detailed spectrally resolved equilibrium and non-equilibrium
THz conductivity dynamics will be presented elsewhere.
\bibitem {comment}To estimate the amount of energy in the high frequency
($>2\Delta$) phonon subsystem, which is in the bottleneck case in the quasi -
thermal equilibrium with the electronic subsystem, we evaluate the energy
required to heat up the high frequency phonon subsystem to $T_{c}$. This
energy in negligible compared to $E_{opt}$ in both NbN and LSCO.
\end{thebibliography}
\end{document}